\documentstyle[12pt,epsfig]{article}

\newcommand{\be}{\begin{equation}}
\newcommand{\ee}{\end{equation}}
\newcommand{\rp}{\right)}
\newcommand{\lp}{\left(}
\newcommand{\bea}{\begin{eqnarray}}
\newcommand{\eea}{\end{eqnarray}}
\newcommand{\eq}{equation}

\begin{document}

\title{Modeling the Stock Market\\ prior to large crashes}
\thispagestyle{empty}

\author{Anders Johansen$^1$ and Didier Sornette$^{1,2,3}$\\
$^1$ Institute of Geophysics and
Planetary Physics\\ University of California, Los Angeles, California 90095\\
$^2$ Department of Earth and Space Science\\
University of California, Los Angeles, California 90095\\
$^3$ Laboratoire de Physique de la Mati\`{e}re Condens\'{e}e\\ CNRS UMR6622 and
Universit\'{e} de Nice-Sophia Antipolis\\ B.P. 71, Parc
Valrose, 06108 Nice Cedex 2, France}

\thispagestyle{empty}

\maketitle

\vskip 2cm

\begin{abstract}

We propose that the minimal requirements for a model of stock market price
fluctuations
should comprise time asymmetry, robustness with respect to connectivity
between agents,
``bounded rationality'' and a probabilistic description. We also compare
extensively
two previously proposed models of log-periodic behavior
of the stock market index prior to a large crash. We find that the model
which follows
the above requirements outperforms the other with a high statistical
significance.

\end{abstract}

{\bf Pacs numbers: 01.75+m ; 02.50-r ; 89.90+n}

\newpage

\pagenumbering{arabic}

\section{General guidelines for stock market modeling}

\noindent {\it Self-organization}\,: Recently, statistical evidence has
shown that
the largest stock market crashes are outliers \cite{JS98.1}. We have
proposed that they have a different origin than the usual day-to-day
variations of the stock market. Several groups have argued that these crashes
have strong analogies to the critical points much studied in statistical
physics [2-13]. The analogy is based on a large body of work exploiting the
many similarities between statistical physics and the financial markets
[14-17].

On what one might refer to as the ``microscopic'' level, the stock market has
characteristics with strong analogies to well-known microscopic models in
statistical physics.
The individual trader has only $3$ possible actions (or ``states''): selling,
buying or waiting. The transformation from one of these states to another
is furthermore a discontinuous process due to a threshold being exceeded,
usually the price of the stock. The transition involves another trader
and the process is irreversible, since a trader cannot sell the stock
he or she bought back to the same trader at the same price. Furthermore, the
individual traders only have information on the action of a limited number
of other traders and in general only see the cooperative response of the
market as a whole in terms of an increase or decrease in the value of the
market.

Strong positive feed-back is also present and is usually referred to
as trend-chasing. Even though one usually divides the market actors into two
classes, typically referred to as fundamentalists\footnote{Fundamentalists
are traders, who base their expectations of the future stock value on
economical criteria.} and technical analysts, all participants
must behave as trend-chasers to some extent in order to maximize
profits\footnote{An investor, who is continuously acting against the trend,
will eventually lose his or her money.}. This means that we are dealing with
a system, where the actions of the trader(s) determine the value of the
stock, which determines the actions of other traders and so forth. Hence, the
stock market is a prime example of a self-organising system and it is thus
natural to think of the stock market in terms complex systems with analogies
to dynamically driven out-of-equilibrium systems such as earthquakes,
avalanches, crack propagation {\it etc.}

\vskip 0.5cm
\noindent {\it Market structure}\,: The stock market also differs
from most spatially extended systems in  several ways. First, there is the
question of ``distances'',{\it i.e.}, when are two traders ``close enough''
to be influencing each other's actions. As the present computerization of
the trading clearly illustrates, this has certainly nothing to do with
the $2-d$ Euclidean space we inhabit and also previously the telephone,
telegraph, Telex {\it etc.} made physical distances of minor importance.
Furthermore, traders do not hold a
``fixed position'' in relation to other traders, but ``move around''
continuously interacting with other traders establishing new correlations.
Second, there is the question of what is the effective drive of the stock
market. Naturally, developments in the world economy play a significant role
in determine the  behaviour of the stock market  (just as the motion of
continental and sub-continental plates due to the heat-transfer from the
interior of the earth plays a role for earthquakes ), but other factors,
{\it e.g.}, to what degree the relevant information is available or the present
``signal-to-noise ratio'', play just as large or even larger role. The
fundamental question is that since we are dealing with a social science, we
enter a field that lacks ``first principles'' and where the fundamental
equations are unknown. This has profound consequences and means that one must
be very  cautious in the choice of approach to the subject. Interesting enough,
we may again turn to the field of statistical physics for inspiration and
guidelines.

\vskip 0.5cm
\noindent {\it Landau expansion}\,: A very powerful and general tool used
in the studies of
phase transitions is that of Landau expansions. In essence, Landau expansions
amounts to assuming some functional relationship $F\lp x \rp$ between the
relevant observable and the corresponding governing parameter. A general form
of an evolution equation for $F$ is can then be derived by expanding $F$
around $0$ to arbitrary order \cite{SJ97}
\be \label{Landau}
\frac{d\log F\lp x\rp}{d\log x}=\alpha F\lp x\rp +\beta  F^2\lp x\rp \ldots ,
\ee
where in general the coefficients may be complex. Using symmetry arguments, one
may reduce the complexity of equation (\ref{Landau}) by proving that
certain coefficients in the expansion must be zero. Furthermore, conservation
rules and equivalent arguments may further simplify the problem putting some
bounds on the values of the remaining coefficients. Since this method only
uses symmetry and conservation arguments, which are independent of the specific
microscopic rules of the system, Landau expansion carries enough generality to
be applied to  a complex problem such as the stock market.

\vskip 0.5cm
\noindent {\it Rationality}\,:
Another basic principle which must be recognized in any modeling approach of
the stock market is that of rationality. Contrary to the general perception in
the physics community of the stock market as populated by irrational herds,
traders in general do exhibit a rational behaviour where they try to optimize
their strategies based on the available information. This one may refer to
as ``bounded rationality'' \cite{boundedratio} since not only is the
available information in
general incomplete, but stock market traders do also have limited abilities
with respect to analysing the available information. This means that the
process of decision making is essentially a ``noisy process'' and, as a
consequence,  that a probabilistic approach in stock market modeling is
unavoidable. Clearly, a noise free stock market with all information
available occupied by fully rational traders of infinite analysis abilities
would have a very small trading volume, if any.

\vskip 0.5cm
\noindent {\it Statistical time asymmetry}\,:
A much neglected question in the modeling of the stock market is that of
symmetry. Most models of the stock market, statistical, such as the GARCH
model \cite{garchref}, or microscopical \cite{StaufferSex,kertesz}
are symmetric with respect to draw-downs
and draw-ups {\it i.e.}, that large/small price movements in either
direction are typically followed by large/small price movements in either
direction. The point we would like to stress here is that this symmetric
behaviour of these artificial indices is not compatible with what is seen
in the Dow Jones Average in general \cite{JS98.1,LeBaron}. Especially in the
case of the large crashes analysed here, it is clear by just looking at the
index that the build-up has been relatively slow and the crash quite rapid.
In other words, ``bubbles'' are slow and crashes are fast. Furthermore,
symmetry with respect to draw-downs and draw-ups is equivalent to a stock
market
dynamics which is invariant with respect to time reversal. That this should
be the case is clearly absurd on any longer time scales\footnote{Since any
long term change is an accumulation of short term changes, symmetry of
short term changes can only be a first approximation.}. This means that not
only must a long-term model
of the stock market prior to large crashes be highly non-linear but it must
also have a ``time-direction'' in the sense that the crash is ``attractive''
prior to the crash and ``repulsive'' after the crash. The importance and
evidence of statistical time reversal symmetry breaking in the sense of
Pomeau \cite{Pomeau} is being increasingly studied \cite{Ramsey}.

\vskip 0.5cm
\noindent {\it Guidelines}\,:
Before we continue with the more technical aspects of our stock market
analysis, let us list what we believe to be essential guidelines in
the modeling of the financial markets:
\begin{itemize}

\item Since the fundamental equations governing the financial markets are
unknown, it is essential that we {\it a priori} identify the symmetry and
conservations laws that applies, if any.

\item Since the participants of the financial markets are unable to fully
optimize their strategies due to incomplete information and limited abilities
with respect to analysing the available information, a probabilistic approach
to stock market modeling is unavoidable.

\item Without systematic comparison between the predictions of the model and
real data, no validation is possible.

\item Any long-term model of the stock market must address the time asymmetry
of market fluctuations observed in the real stock market.

\end{itemize}

In the next section we will briefly describe a model which is based on these
fundamental guidelines. In the third section, we compare the predictions of
our model with real stock market data and briefly discuss an alternative
approach to such data estimation. The last section concludes.

\section{The model}

The model is explained in detail in \cite{JLS} and we briefly
list the key-assumptions and basic components.
The first is that a large crash is caused by {\em local} self-reinforcing
imitation between traders. In the presence of noise, this self-reinforcing
imitation process eventually leads to a bubble. If the tendency for traders
to ``imitate'' their ``friends'' increases up to a certain point called the
``critical'' point, many traders may place the same  order (sell) at the same
time, thus causing a  crash.

The interplay between the progressive
strengthening of imitation and the ubiquity of noise requires a stochastic
description. This means that a crash is not certain but that its probability
can be characterized
by a rate $h(t)$ equal to the probability per unit time that the crash will
happen in the next instant if it has not happened yet.

Second, since the crash
is not a certain deterministic outcome of the bubble, it remains rational for
traders to remain invested provided they are compensated by a higher rate of
growth of the bubble for taking the risk of a crash, because there is a finite
probability of  ``landing'' smoothly, {\it i.e.} of attaining the end of the
bubble without crash. In this model, the ability to predict the critical date
is perfectly consistent with the behavior of the rational agents: they all
know this date, the crash may happen anyway, and they are unable to make any
abnormal risk-adjusted profits by using this information. We emphasize that
the model
distinguishes between the end of the bubble and the time of the crash\,: the
rational expectation constraint has the specific implication that the date of
the crash must have some degree of randomness. Hence, the theoretical death of
the bubble is not the time of the crash and the crash could happen at any time
before, however not very likely. The death of the bubble is simply the most
probable time for the crash.

The model does not impose any constraint on the amplitude of the crash. We
have considered two possibilities, which offers a plausible scenario for
bubbles on long and short time scales, respectively. If we assume that the
size of the crash is proportional to the current price level, then the natural
variable is the logarithm of the price. This is the hypothesis which was
pursued in \cite{JLS} considering build-ups over $\approx 8$ years. If instead,
we assume that the crash amplitude is a finite fraction of the gain observed
during the bubble, then the natural variable is the price itself. This
hypothesis should apply to relatively short time scales of about two years and
was pursued in \cite{JS98.2}.

A crash happens when a large group of agents place sell orders simultaneously,
which brings us to the question of how the agents interact. We have proposed
the following description \cite{JLS}: all the traders in the world are
organised
into networks of family, friends, colleagues, etc. Hence, the opinion of a
trader is influenced by (a) the opinions of these people and (b) an
idiosyncratic signal that the trader alone generates.

The last ingredient of the model is to recognize that the stock market is made
of actors which differ in size by many orders of magnitudes ranging from
individuals to gigantic professional investors, such as pension funds.
Structures at even higher levels, such as currency influence spheres (US\$,
Euro, YEN ...), exist and with the current globalisation and de-regulation of
the market one may argue that structures on the largest possible scale,
{\it i.e.}, the world economy, are beginning to form. This means that the
structure of the financial markets have features which resemble that of
hierarchical systems with ``agents'' on all levels of the market. (Of course,
this does not imply that any strict hierarchical structure of the stock market
exists.) Models \cite{JLS,SJ98} of imitative interactions on hierarchical
structures predicts that the first order expansion of the general solution for
the crash hazard rate is then
\be
\label{eq:hazard3}
h(t)\approx B_0(t_c-t)^{-\alpha}
+B_1(t_c-t)^{-\alpha}\cos[\omega\log(t_c-t)-\psi']
\ee
and similarly that the evolution of the
price  before the crash and before the critical date is given by:
\be
\label{eq:complex}
p(t) \approx p_c -\frac{\kappa}{\beta}\left\{
B_0(t_c-t)^{\beta}
+B_1(t_c-t)^{\beta}\cos[\omega\log(t_c-t)-\phi]\right\}
\ee
where $\phi$ is another phase constant. The key feature is that log-periodic
oscillations appear in the price of the asset before the critical date $t_c$.
These oscillations are controlled by a prefered scaling ratio $\lambda =
e^{2\pi/\omega}$ characterising the hierarchical structure of the stock market.
Including the next order term in the expansion is useful for analyzing data
over long period of times \cite{SJ97}
for which the relevant observable is the logarithm of the price\,:
\be
\label{eq:nonlinear}
\log\left[\frac{p(t)}{p_c}\right]\approx
\frac{(t_c-t)^\beta}{
\sqrt{1+\left(\frac{t_c-t}{\Delta_t}\right)^{2\beta}}}
\left\{B_0+B_1\cos\left[\omega\log(t_c-t)+\frac{\Delta_\omega}{2\beta}
\log\left(1+\left(\frac{t_c-
t}{\Delta_t}\right)^{2\beta}\right)-\phi\right]\right\} .
\ee

\section{Data analysis}

\subsection{Previous results}

In a series of work \cite{SJB96,FF96,SJ97,THESIS,JS98.2,van1} it has been shown
that the time-evolution of the stock market index prior to the 1929, 1987
and 1998 crashes on Wall Street and the 1997 crash in Hong-Kong are in very
good agreement with the predictions of the model presented in the previous
section. Specifically, the Dow Jones and the S\&P 500 was well-described
\cite{JS98.2} by equation (\ref{eq:complex}) over a time interval
approximately $2.5$ years
prior to the four crashes mentioned. Furthermore, for the 1929 and 1987
crashes on Wall Street
this time interval could be extended to almost $8$ years using equation
(\ref{eq:nonlinear}).

As discussed in the previous section, within the framework
of the model presented here as well as from general arguments, the exponent in
equations (\ref{eq:complex}) and (\ref{eq:nonlinear}) must obey the
inequalities
$0 < \beta < 1$. The first inequality ensures that the stock market index
remains finite. The second inequality describes an acceleration of the bubble
and of the crash hazard rate.
In the case of the four crashes mentioned above, we consistently
found \cite{JS98.2} values for $\omega$ in the range $\left[ 6.4:7.9\right]$
corresponding to a preferred scaling ratio of $\lambda = e^{2\pi /\omega} =
2.45 \pm 0.25$ governing the log-periodic oscillations. For the exponent
we found a somewhat broader range of values with $\beta_{29} = 0.45$,
$\beta_{87} = 0.33$,
$\beta_{97} = 0.34$ and $\beta_{98} = 0.60$. That the values of the
preferred scaling ratio is within $10\%$ for the four large crashes in this
century is in fact
quite amazing {\it a priori} and reassuring from a theoretical point of view.
Also the larger fluctuations in the value of the exponent can be explained
within the proposed framework, since it is known that noise will
renormalize the exponent. In Ref.\cite{SS}, it is shown that the exponent
$\beta$ is realization dependent and obeys the following equation\,:
\be
\lambda^{\beta} = \mu~,~~~~~~{\rm leading~to}~~\beta = {1 \over 2\pi}
~\omega~\ln \mu~,
\label{dfghnwxw}
\ee
where $\lambda = e^{2\pi / \omega}$. The parameter $\mu$ is a scaling
factor for the
observable. Expression (\ref{dfghnwxw}) shows that realization dependent
fluctuations of
$\omega$ and $\ln \mu$ contribute multiplicatively to the fluctuations of
$\beta$.
If the relative magnitude of the fluctuations ${\Delta \ln \mu \over \ln
\mu}$ of $\ln \mu$
is of the same order as ${\Delta \omega \over \omega} = {\Delta \lambda
\over \lambda}\approx 10\%$,
we obtain the
estimation ${\Delta \beta \over \beta} \approx 20\%$, which may rationalize
the
range of values $0.33-0.45$ found for $\beta$ for three of the crashes at
the exclusion
of the most recent one in August 1998. A possible explanation for the
anomalously large value $\beta_{98} = 0.60$ is the ``grey-monday''
Oct. 1997 --$7\%$-correction-- on Wall Street, which clearly decreased
the acceleration of the index.

These fluctuations in the exponent $\beta$ have led
other groups to suggest a different scenario than the one proposed by equation
(\ref{eq:complex}). In two recent papers \cite{van1,van2} N. Vandewalle
{\it al.} suggest that equation (\ref{eq:complex}) should be replaced with
\bea \label{logeq}
F\lp t\rp &=& A + B\ln\lp \frac{t_c -t}{t_c}\rp \left[ 1 +
C\cos\lp \omega \ln \lp \frac{t_c -t}{t_c}\rp + \phi \rp \right] .
\eea
Equation (\ref{logeq}) is well-known in statistical physics where it describes
a  special class of phase transitions characterized by a logarithmic
divergence.

There are a number of implications involved in replacing equation
(\ref{eq:complex})
with equation (\ref{logeq}) not explicitly considered by N. Vandewalle
{\it al}.
In their papers, they state that \eq \ (\ref{logeq}) correspond to equation
(\ref{eq:complex}) in the limit $\beta \rightarrow 0$. This is not entirely
true, which can
easily be seen by rewriting the leading term of \eq \ (\ref{eq:complex})
substituting
$x=t/t_c$
\bea
p\lp t\rp &=& A + B\lp 1-x\rp^\beta =A+Be^{\beta\ln\lp 1 - x\rp}
\\ \label{1fexpan}
 &\approx& A'+B' \ln\lp 1 - x\rp \mbox{\ for \ } \beta\ln\lp 1 - x\rp \ll 1
\eea
Hence the expansion above is valid in the limit $\beta \rightarrow 0$
{\it provided} the  term $\ln\lp 1 - x\rp$ remains finite. This touches
with a rather troublesome feature of \eq \ (\ref{logeq}). Going from equation
(\ref{eq:complex}) to
equation (\ref{logeq}) implies that the value of the stock market index at the
time of the crash $t_c$ is no longer finite but diverges, {\it i.e.}, in the
two  cases we have $\lim_{t\rightarrow t_c}F\lp t\rp =A$ and
$\lim_{t\rightarrow
 t_c} F\lp t \rp =\infty$, respectively\footnote{N. Vandewalle {\it al.}
report
incorrectly the sign of the exponent in their reference of our previous work
and hence erroneously conclude that the proposed relation
(\protect\ref{eq:complex})
diverges for $t=t_c$. Let us also note that \eq \ (3) in \protect\cite{van2}
should
read $\frac{t_{n+1}-t_n}{t_n - t_{n-1}} = \lambda$ which is an example
of Shank's transformation for the acceleration of the convergence approximately
geometric series \cite{Benderor,Revue}. Last, the correct relation
between $\lambda$ and $\omega$ is $\omega \ln \lambda = 2\pi$ and not
$\ln \lambda = \omega/2\pi$ as reported in \protect\cite{van2}.}. The notion
of the stock market index going to infinity seems rather unrealistic for the
following reasons. A stock market with infinite prices means that all traders
must have infinite wealth, not only one. If we assume steady cash-flow and not
hyper-inflation this is clearly impossible. Furthermore, if we believe 
Pareto's law with its finite first moment to be a valid discription of wealth 
distribution, prices cannot go to infinity.

Another theoretical objection to \eq \ (\ref{logeq}) comes from the
fact that we are considering a very special class of models with logarithmic
divergences. If, as we believe, phase transitions provides
a valid framework for the description of certain dynamical features in the
financial markets, it seems very difficult to justify that the analogies
between phase transitions and large stock market crashes should be limited
to a very special class characterized by logarithmic divergens.  On the
contrary, we have above proposed a model which does not make unrealistic
assumptions about the dynamics of the stock market nor
contains economical abnormalities such as an infinite stock market index.

\subsection{Which universality class?}

As previously mentioned, the two equations do have a similar behaviour for
$\beta \ln \lp 1-t/t_c\rp \ll 1$. Hence, from a technical trading point of
view it might be interesting to compare the numerical performance of \eq s
(\ref{eq:complex}) and (\ref{eq:nonlinear}) on one hand and (\ref{logeq})
on the other. Especially, the condition $\beta \ln \lp 1-t/t_c\rp \ll 1$
need to be quantified, {\it i.e.}, when are we too close to $t_c$ for \eq \
(\ref{logeq}) to be a good approximation. In order to do so in a consistent
way,
we have implemented \eq \  (\ref{logeq}) in the same numerical algorithm used
in \cite{SJB96,SJ97,THESIS,JLS} in fitting \eq \ (\ref{eq:complex}) and its
second order extension to the stock market index, \eq \ (\ref{eq:nonlinear}).
We emphasize that contrary to the view expressed in \cite{van2}, the numerical
estimation procedure of fitting \eq  (\ref{eq:complex}) used does not involve
a 7-parameter nonlinear fit. This, since {\it any} linear parameter in a
fitting
function can be expressed in terms of the nonlinear parameters by demanding
that the chosen cost-function has a zero derivative in a minimum with respect
to the linear variables. This means that fitting \eq s (\ref{logeq}),
(\ref{eq:complex}) and (\ref{eq:nonlinear}) simply amounts to a 3-, 4- and
6-parameter nonlinear fit, respectively. Among
these, the phase $\phi$ is simply a time-unit and  has no influence on the
value of the other parameters\footnote{This can easily be verified by a change
in time-unit from, {\it e.g.}, days to months: the value of $\phi$ is
changed, but the value of the other parameters remains the same.}.

Before we
proceed with the data analysis, let us briefly comment on the numerical work
presented in \cite{van2}. The authors claim that one of the advantages of
\eq \ (\ref{logeq}) over \eq \  (\ref{eq:complex}) is that the number of
parameters to fit is reduced by one  {\it yet they do not fit} \eq \
(\ref{logeq}) to the stock market index. Instead they truncate \eq \
(\ref{logeq}) and only fit the leading term setting $C=0$. This fit is
then compared with a pure power law plus a constant, {\it i.e.}, \eq \
(\ref{eq:complex}) with $C=0$. The period of the log-periodic oscillations
is instead estimated by localizing in a non-systematic way, specifically
by eye-balling, the local extrema of the log-periodic oscillations. This
procedure will in general introduce some systematic errors. First, since the
logarithmic law (or for that sake a pure power
law) can only capture the average rise in the data, the fit becomes sensitive
to whether we use a minimum or a maximum as the starting point for the
data set. Specifically, the fit of the S\&P500 with a pure logarithmic rise
gives $t_c=87.96$ using the minimum of $80.0$ as first point and $t_c=88.07 $
using the following maximum of $81.24$ as first point, {\it i.e.} a difference
of $\approx 35$ days. For a much shorter time series the effect is of course
much more dramatic. Second, estimating maxima and minima by eye-balling may be
quite reasonable as a first estimate. However, pattern recognition in general
{\it must}  be based on a deterministic algorithm in order to carry any
real weight.
Furthermore, it is not possible to compare the performance of different
fitting functions if there are not implemented using more or less the same
standard procedures. Needless to say, the eye-balling performed in
\cite{van2,laloux} is not such a standard  procedure.

\begin{figure}
\begin{center}
\epsfig{file=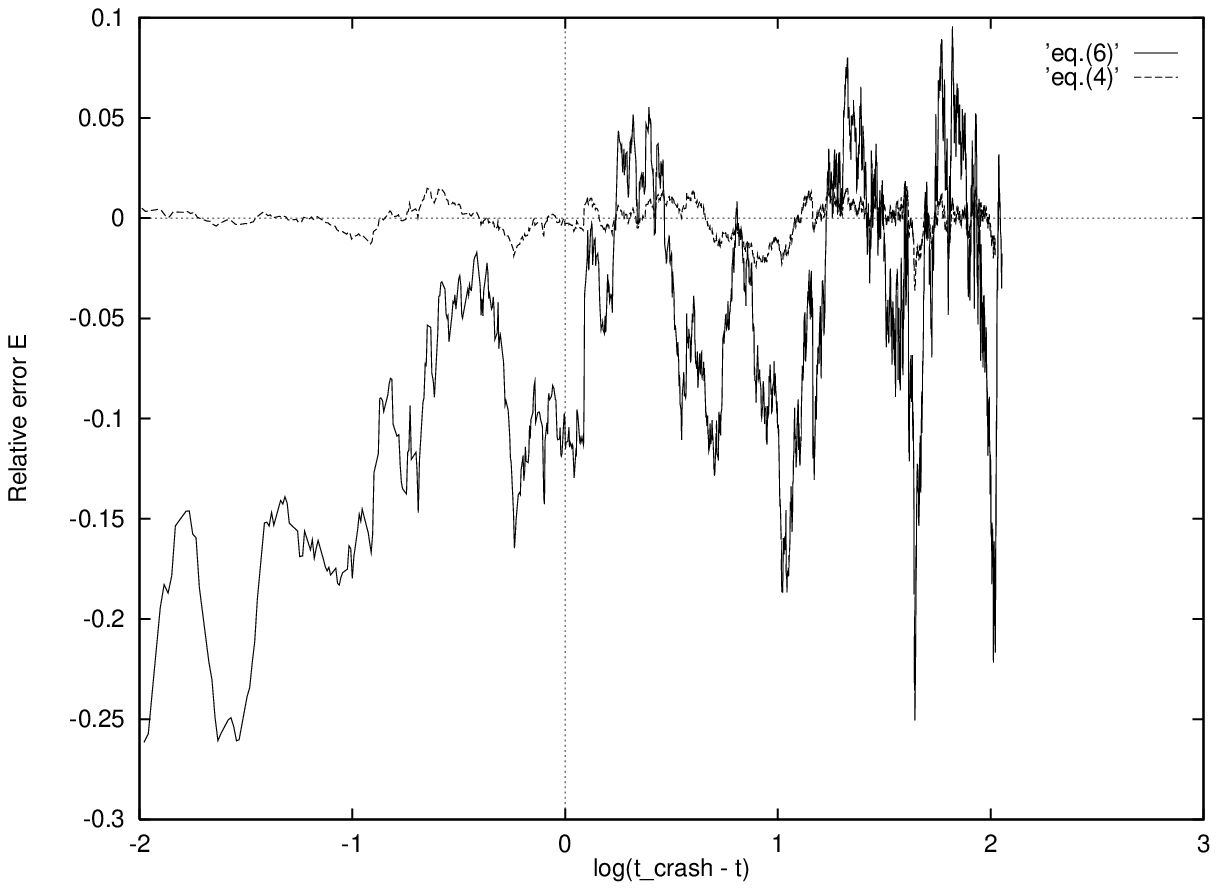, width=12cm,height=10cm}
\caption{\protect\label{johansen1}The relative error $E$, as defined by
(\ref{erroer}),
between the S\&P 500 and \eq \ (\protect\ref{logeq}) and
$\log\lp \mbox{S\&P 500}\rp$ and \eq \ (\protect\ref{eq:nonlinear}), 
respectively.
The time period shown is from $80.24$ to $87.65$ measured in units of
$\ln \lp t_{crash} - t\rp$, where  $t_{crash}=87.79$ is the
time of the crash. The larger fluctuations belong to the fit of
\eq \ (\ref{logeq}). The parameters for the fit of \eq \ (\ref{logeq}) is
$A\approx 263$, $B\approx  -76$, $BC\approx 9.6$, $t_c\approx  88.03$,
$\phi\approx  0.92$ and $\omega\approx 13.4$. The parameters for the fit of
\eq \ (\protect\ref{eq:nonlinear}) is $\log(p_c) \approx 5.90$,
$B_0 \approx -0.38$,
$B_1\approx 0.044$,  $\beta\approx 0.68$, $t_c\approx
87.81$, $\phi\approx  -2.8$, $\omega\approx  8.9$, $\Delta_t\approx  11$ and
$\Delta_\omega\approx 18$.}
\end{center}
\end{figure}

\begin{figure}
\begin{center}
\epsfig{file=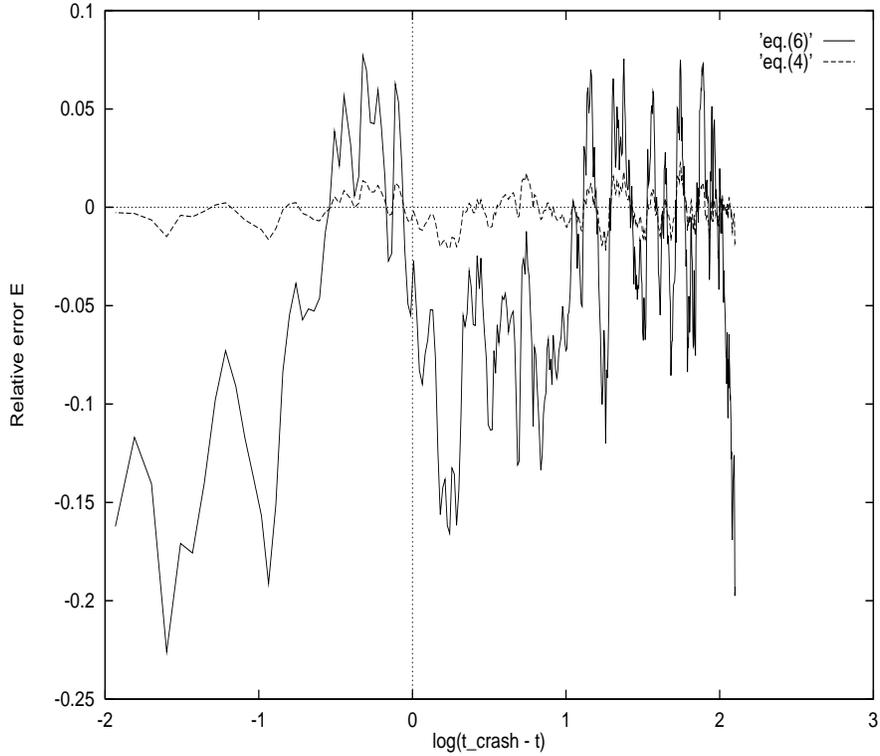, width=12cm,height=10cm}
\caption{\protect\label{johansen2} The relative error $E$, as defined by
(\ref{erroer}),
between the  Dow Jones and \eq \ (\protect\ref{logeq}) and
$\log\lp \mbox{Dow Jones}\rp$ and  \eq \ (\protect\ref{eq:nonlinear}),
 respectively.
The time period shown is from $21.05$ to $29.66$ measured in units of
$\ln \lp t_{crash} - t\rp$, where $ t_{crash}=29.81$ is the time
of the crash. The larger fluctuations belong to the fit of
\eq \ (\ref{logeq}). The parameters of the best fit of \eq \ (\ref{logeq}) is
$A\approx  264$, $B\approx  -92$, $BC\approx  -7.3$, $t_c\approx  29.94$,
$\phi\approx  -2.1$ and $\omega\approx  10.0$. The parameters obtained for
the fit of \eq \ (\protect\ref{eq:nonlinear}) is $\log(p_c)\approx  6.11$,
$
B_0\approx  -0.565$,
$B_1\approx  0.046$, $\beta\approx  0.63$, $t_c\approx
29.84$,  $\phi\approx  3.0$, $\omega\approx  5.0$, $\Delta_t\approx 14$
and $\Delta_\omega\approx  -70.0$.}
\end{center}
\end{figure}

\begin{figure}
\begin{center}
\epsfig{file=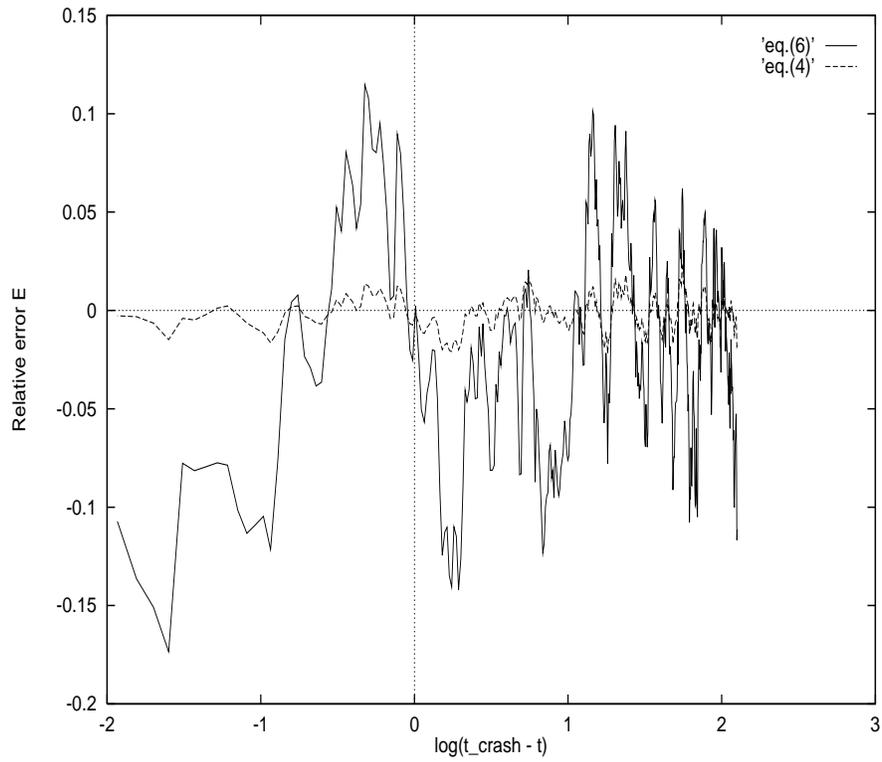, width=12cm,height=10cm}
\caption{\protect\label{johansen3} Same as figure \protect\ref{johansen2} but
for the second best fit of \eq \ (\protect\ref{logeq}). The parameters
for the second best fit of \eq \ (\protect\ref{logeq}) are $A\approx  257$,
$B\approx  -88$, $BC\approx  6.8$, $t_c\approx  29.88$, $\phi\approx  -3.5$
and $\omega\approx  11.0$. }
\end{center}
\end{figure}

\begin{figure}
\begin{center}
\epsfig{file=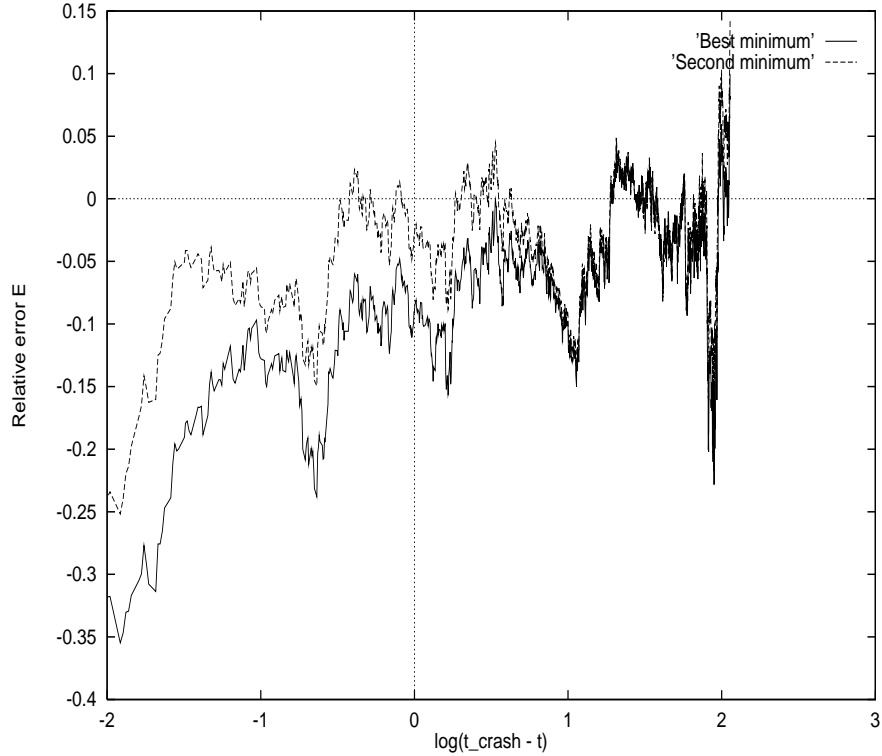, width=12cm,height=10cm}
\caption{\protect\label{johansen4} The relative error $E$, as defined by
(\ref{erroer}),
between the S\&P 500 and the two best fits of \eq  \ (\protect\ref{logeq}).
The time period shown is from $90.01$ to $97.77$ measured in units of
$\ln \lp t_{crash} - t\rp$, where $t_{crash}=97.81$ is the
time of the
crash. The parameters of the best fit of \eq \ (\protect\ref{logeq})
is $A\approx  771$, $B\approx  -224$, $BC\approx  21$, $t_c\approx  98.12$,
$\phi\approx  -4.5$ and $\omega\approx  4.5$. The parameter values of the
second fit is $A\approx  727$, $B\approx  -195$, $BC\approx  16$, $t_c\approx
97.97$, $\phi\approx  -2.5$    and $\omega\approx  5.9$.}
\end{center}
\end{figure}

In order to have a completely deterministic algorithm, we have fitted the
stock market data using the same standard numerical algorithms as in our
previous work (see \cite{THESIS} for details). In order to fully
compare the two approaches, we have chosen two large crashes with an
almost $8$ years build-up as our test sample. Specifically,
we have fitted the index almost 8 years prior to the 1929 and 1987 crashes on
Wall Street with \eq s (\ref{logeq}) and \eq \ (\ref{eq:nonlinear}). In
addition, we have tested the prediction of the ``grey-monday'' Oct.
1997 --$7\%$-correction-- on Wall Street by N. Vandewalle {\it et al.}
 fitting the full
equation (\ref{logeq}) on the time-interval used in \cite{van2}. We have
also compared  \eq \ (\ref{logeq}) with \eq \ (\ref{eq:complex}) on shorter
time intervals of $2-3$ years analysed for four crashes in \cite{JS98.2}.
As indicated by the expansion (\ref{1fexpan}) and shown in figures
\ref{johansen1}-\ref{johansen4}, \eq \ (\ref{logeq}) does not perform very well
``close'' to the crash. Hence, the results from fitting \eq \ (\ref{logeq})
with these shorter time intervals were not convincing at all and we have not
included the analysis on the shorter time intervals in the present paper.

\subsubsection{Fitting stock markets data}

In figures \ref{johansen1}-\ref{johansen4} we see the relative error $E$ of the
fitting function $f$ to the data $y$ defined as
\be
E\lp t\rp = \frac{y\lp t\rp - f\lp t \rp}{y\lp t\rp}
\label{erroer}
\ee
as a function of $\log \lp t_{crash} - t\rp$ for the 1929 and 1987 crashes on
Wall Street. The interval fitted starts almost $8$ years  prior to the two
crashes and ends at the date where the index has achieved its maximum. In order
to compare the fits of \eq s (\ref{logeq}) and (\ref{eq:nonlinear}), we have
used $t_c = t_{crash}$ in estimating the relative error of the fits to
the data. Specifically, the dates used as $t_{crash}$ for the three
data sets was $29.81$, $87.79$ and $97.81$,

In the case of the 1987 crash, the best fit also gave the most accurate
estimate for $t_{crash}$ for both formulas. In the case of the 1929
crash, we show the two best fit of \eq \ (\ref{logeq}) together with the best
fit of \eq \ (\ref{eq:nonlinear}) according to the criterions described in
\cite{SJ97,THESIS,JLS}. A few things are worth stressing:

\begin{itemize}

\item  The fit with equation (\ref{logeq}) overshoots $t_{crash}$
significantly more than reported in \cite{van2}.

\item  The values obtained for $\lambda$ fitting equation (\ref{logeq})
fluctuates by $\pm 40\%$, which is considerably more
than the $\pm 10\%$ that is experienced with (\ref{eq:complex}) and
\eq \ (\ref{eq:nonlinear}) \footnote{The $40\%$ is also
considerably larger than found in \cite{van2} using eye-balling.}.

\item  Equation (\ref{logeq}) does not seem to capture the log-periodic
oscillations very effectively.

\item The error between the data and equation (\ref{logeq}) grows
significantly as $t_{crash}$ is approached.

\end{itemize}

What is interesting in this comparison is not so much the larger error by
\eq \ (\ref{logeq}) compared to \eq \ (\ref{eq:nonlinear}) for the 1929
and 1987 crashes, since this is to be expected due to the larger number of
free parameters in the fit (3 against 6). The really interesting difference
between the two approaches are the last two items above, and especially the
fact that the error start to grow around nine month prior to $t_{crash}$
or earlier. This clearly shows that \eq \ (\ref{logeq}) is {\it only} a good
approximation far from $t_{crash}$. The only possible conclusion in our 
opinion is that the time-dependent acceleration of the market price is 
{\it not} logarithmic and that the range where expansion (\ref{1fexpan})
is valid is surprisingly restricted.

\section{Conclusion}

We have proposed a general set of simple guidelines for the
modeling of financial markets. Using these guidelines, we have highlighted the
key ingredients of
a rational expectation model of the stock market. We find an excellent
agreement between the predictions of the model and the evolution of the
Dow Jones and S\&P 500 indices prior to the largest crashes of this century.
Last, we have shown that the proposition of N. Vandewalle {\it al.} of
stock market crashes not only leads to economical abnormalities but also do 
not perform well when compared to data.

\vskip 0.5cm
\noindent {\bf Acknowledgment}

\noindent The authors are grateful to O. Ledoit and Dietrich Stauffer
for many stimulating discussions.

\end{document}